\documentclass[12pt,aps,preprint,floatfix,nofootinbib]{revtex4}

\usepackage{graphicx}
\usepackage{dcolumn}
\usepackage{bm}

\def\beq{\begin{equation}}
\def\eeq{\end{equation}}
\def\bea{\begin{eqnarray}}
\def\eea{\end{eqnarray}}

\usepackage{color}

\begin{document}

\preprint{IPMU09-0130}
\preprint{UCI-TR-2009-18}

\title{Explorations of the Top Quark Forward-Backward Asymmetry\\ at the Tevatron}

\author{Jing Shu~$^{a}$}\email{jing.shu@ipmu.jp}
\author{Tim M.P. Tait~$^{b}$ }\email{ttait@uci.edu}
\author{Kai Wang~$^{a}$}\email{kai.wang@ipmu.jp}
\affiliation{$^{a}$ Institute for the Physics and Mathematics of the Universe, The University of Tokyo, Chiba $277-8568$, Japan\\
 $^{b}$ Department of Physics and Astronomy, University of California, Irvine, CA 92697}

\date{\today}
             
\begin{abstract}
We consider the recent measurement of the top quark forward-backward asymmetry at
the Fermilab Tevatron, which shows a discrepancy of slightly more than 2$\sigma$ compared
to the SM prediction.  We find that $t$-channel exchange of a color sextet or triplet scalar
particle can explain the measurement, while leaving the cross section for $t \bar{t}$ production
within measured uncertainties.  Such particles have good discovery prospects by study of
the kinematic structure of $t \bar{t}$+jets at the LHC.
\end{abstract}


\maketitle

\section{Introduction}

The huge mass of the top quark, the only Standard Model (SM) fermion whose mass lies at the
electroweak scale, may be a clue that the top is special in some way, perhaps
serving as a portal to the physics of electroweak symmetry-breaking.  Thanks to the wealth of data from the Tevatron, the existence of top is well-established, and fundamental measurements such as its mass have become routine. Interest now turns to more subtle properties, including the characteristics of its production.  

Recently, both the CDF and D0 collaborations have measured the 
forward-backward asymmetry of the top quark,
$A_{FB}^t$ \cite{cdf,newcdf,d0}.  The most recent
is by CDF based on a data sample of $3.2$~fb$^{-1}$~\cite{newcdf},
\bea
A_{FB}^t  &=& 0.193 \pm 0.065_{\, \rm stat.} 
\pm 0.024_{\, \rm syst.}
\label{eq:newcdf}
\eea
in the $p \bar{p}$ rest frame (with $m_t = 175$ GeV). Combined in quadrature, the over-all measurement is
$A_{FB}^t = 0.193 \pm 0.069$.  This is to be compared with the best theoretical
predictions for $A_{FB}^t$ in the SM~\cite{Antunano:2007da,Bowen:2005ap,mynlo},
\bea
A_{FB}^{t~(SM)} &=& 0.051 ,
\label{eq:SM}
\eea
which in the SM is dominantly produced by one-loop corrections from the strong force,
with a smaller contribution from electroweak $t \bar{t}$ production.
It further appears to be stable with respect to corrections from QCD threshold 
resummation~\cite{Almeida:2008ug}.

This is an interesting measurement for a number of reasons.  First, and most obviously, the
measurement is slightly more than $2 \sigma$ away from the SM prediction.  While the discrepancy
could be simply statistics, it is large enough, and consistent (within reasonably large error bars) 
with the D0 and previous CDF measurements.   Second, this discrepancy is the latest in
a series of measurements of heavy quark asymmetries (starting with $A_{FB}^b$ and 
$A_{FB}^c$ at LEP and SLC) to show a discrepancy, 
raising the question: is this an unrelated effect, or part of a bigger picture?
Finally, this measurement involves the heavy top quark and at very high momentum transfer,
both of which suggest it may be particularly sensitive to physics beyond the Standard Model.

Such a large enhancement of $A_{FB}^t$ is a challenge for physics beyond the Standard
Model.  Measurements of the $t \bar{t}$ inclusive cross section \cite{cdf-ttbar,Abazov:2009ae},
currently dominated by the 4.6~fb$^{-1}$ CDF result ($m_t = 172.5$ GeV),
\bea
\sigma_{t \bar{t}} &= & 7.50 \pm 0.31_{\mathrm{stat}} \pm 0.34_{\mathrm{syst}} \pm 0.15 _{\mathrm{th} } ~ \mathrm{pb}  \ ,
\eea
(combining errors in quadrature, $\sigma^{exp} = 7.50 \pm 0.48 $ pb)
is in agreement with SM theory predictions of $\sigma = 7.5^{+0.5}_{-0.7}$
\cite{Cacciari:2008zb,Kidonakis:2008mu,Moch:2008ai}. 

The $t \bar{t}$ invariant mass distribution appears to fall off as expected
for large invariant masses \cite{Aaltonen:2009iz,D0:dsdmtt}.
Any theory which attempts to "fix" the value of $A_{FB}^t$ must do so without introducing large
corrections to either the cross section or the invariant mass distribution which are inconsistent
with those measurements. 

In Ref.~\cite{Jung:2009jz}, it was argued that a $t$-channel vector 
boson exchange with flavor-violating couplings to right-handed up-type quarks can satisfy all of
these criteria. A $t$-channel exchange avoids resonant behavior, and thus does not lead to large
features in the invariant mass distribution for a light particle, and
allows a relatively larger forward-backward asymmetry compared to $s$-channel exchange \cite{Rodrigo:2008qe, Ferrario:2009bz, Ferrario:2008wm, Martynov:2009en, Djouadi:2009nb, Frampton:2009rk}. In this article, we explore new scalar bosons (in a variety of $SU(3)_c$ representations) which couple in a flavor-violating way to up quarks
and the top quark, and explore the possibility that they can explain the measurement of
$A_{FB}^t$ while remaining consistent with all other measurements. 

\section{Effective Theories}
\label{sec:model}

The SM $SU(3)_c \times SU(2)_L \times U(1)_Y$ gauge structure severely limits the
possible representations of any scalar particle which can couple an up quark
to the top quark.  The various possibilities may be distinguished by
the $SU(3)_c$ representation of $\phi$.  The possible cases include
a color octet, a color singlet, a color triplet, and a color sextet,
\bea
(8, 2)_{-1/2},  ~~(1, 2)_{-1/2}, ~~
(\bar{3}, 1)_{4/3}, ~~ (6, 1)_{4/3}~.
\label{scalar}
\eea
All of these possibilities require that $\phi$ is complex.  We could also explore
$(6,3)_{1/3}$ and $(\bar{3},3)_{1/3}$ representations, but these objects have couplings 
to left-handed
quarks which are (somewhat) related to CKM elements, and thus have more
potential to lead to strong constraints from flavor observables.

To describe the scalar, we add terms (after electroweak symmetry-breaking) 
to the SM Lagrangian such as,
\bea
\mathcal{L}_\phi & = &
D_\mu \phi^\dagger D^\mu \phi - M_\phi^2 | \phi|^2  +
\phi^a \bar{\hat{t}} T_r^a (y_S + y_P \gamma_5) u  + h.c.  , 
\eea
where $D_\mu \phi$ is the appropriate covariant derivative for $\phi$ depending on its
gauge quantum numbers, $M_\phi$ is the scalar mass
and $T_r^a$ are the $SU(3)_c$ Clebsch-Gordon coefficients which connect 
$\phi^a$ of color $a$ to the two quarks.  For the cases in which
$\phi$ lives in an $SU(2)$ doublet, there will also be couplings
between the $SU(2)$ charged partners of the neutral $\phi$ involving the $b$
and $d$ quarks.
$\hat{t}$ depends on the color
representation of $\phi$,
\bea
\hat{t} & = & \left\{
\begin{array}{lc} 
t & {\rm (octet~or~singlet)} \\
t^c  & {\rm (triplet~or~sextet)}
\end{array}
\right.
\eea 
where $t^c = i \gamma^0  \gamma^2 t$ is the charge conjugate of the top quark.

We have neglected the possibility of a quartic interaction for $\phi$
(or mixed $\phi$ Higgs quartic interactions), full $3 \times 3$ generational
coupling to the up-type quarks, and coupling to down-type quarks.  All of these
may be included without
substantially changing our conclusions, although the down-type couplings will in general
have strong experimental constraints from flavor-violating observables.

\section{Asymmetry and Cross Section}

The process $ u (p_1) \; \bar u (p_2) \to t (k_1) \; \bar t(k_2)$ is 
described in the SM by the $s$-channel gluon exchange diagram.
In additional, the flavor-violating portion of the $\phi$ interaction will
mediate a $t$-channel contribution.  The differential 
partonic cross section, summed/averaged over 
final/initial state spins and colors may be written,
\begin{equation}
 \frac{ d \sigma}{ d \cos\theta} = \frac{1}{4} \frac{1}{9} \frac{\beta} {32 \pi s}\,
 \sum \left| { {\cal M} }  \right |^2 \;,
\end{equation}
where $\theta$ and
 $\beta \equiv \sqrt{1 -  4 m^2_t / s}$ are the  the scattering angle between the outgoing top and
 incoming quark and
 top quark velocity, defined in the partonic center-of-mass frame,
and the sum is over the matrix elements ${\cal M}$ for both SM and $\phi$ contributions.
The hadronic cross section is obtained by convolving this cross section with the parton
distribution functions (PDFs).

Including both the SM and $\phi$ contributions,
\begin{eqnarray}
\label{eq: scalar_stu}
 \sum  \left | {\cal M} \right |^2  &=& \frac{16 g_S^4}{s^2} (t_t^2 +u_t^2 +2 s m_t^2)  
  + 8 C_{(0)} g_S^2 y^2 \frac{ s m_t^2 +t_t^2}{s t_\phi}
+ C_{(2)} 
\frac{4 y^4 t_t^2} {t_\phi^2} 
\ ,
\end{eqnarray}
where the tiny up quark mass has been neglected, and
$s \equiv (p_1 + p_2)^2$, 
$t \equiv (p_1 - k_1)^2$, and
$u \equiv (p_1 - k_2)^2$ are the Mandelstam variables.  We use a compact notation where
$t_t \equiv t - m_t^2$, $t_\phi \equiv t - M_{\phi}^2$, and so on.  The couplings $y_S$ and
$y_P$ occur only in the combination $y \equiv  \sqrt{y_S^2 + y_P^2}$, indicating that at this level
our results are the same regardless of the chiral nature of the $\phi$ coupling.  The color factors
$C^{(0)}$ and $C^{(2)}$ depend on the color representation of $\phi$.  They can be found in 
Table~\ref{tab:color}.  The color sextet and triplet cases also require the switch of
Mandelstam $u \leftrightarrow t$.

\begin{table}[!tb]
\begin{tabular}{| c | c c c c |}
\hline
 ~~Color Factor~~     &   ~~Octet~~  &  ~~Singlet~~ &    ~~Sextet~~  &    ~~Triplet~~  \\
 \hline
 $C_{(0)}$    &   $-2/3$ & $4$ & $1$ & 1\\
 $C_{(2)}$  &   $2$ & $9$ & $3/2$ & 3/4 \\
\hline
\end{tabular}
\caption{Color factors for a $SU(3)$ octet, singlet, sextet, and triplet object in the $t$-channel. $C_{(0)}$ and $C_{(2)}$ stand for the color factor from the $t$-channel physics interfering
with an $s$-channel gluon exchange and its amplitude squared term, respectively. }
\label{tab:color}
\end{table}

The dependence on the scattering angle, and thus the potential to contribute 
to a forward-backward
asymmetry, may be made more clear by working in the partonic center-of-mass frame variables
$\theta$ and $\beta$ defined above.  Using the relations, $u_t = -s(1+ c_\theta)/2$, 
$t_t = -s(1- c_\theta)/2$, where $c_\theta = \beta \cos {\theta}$ for a color octet or
singlet $\phi$ (for a sextet or triplet, the swap $t \leftrightarrow u$ takes 
$c_\theta \rightarrow - c_\theta$), the matrix element becomes,
\begin{eqnarray}
\label{eq: scalar_cos}
& & \hspace*{-0.65cm} \sum  \left | {\cal M} \right |^2
  = 8 g_S^4 (1+c_\theta^2+4m^2 )  + 
2 y^2 g_S^2 C_{(0)}  s \frac{ (1-c_\theta)^2 +4m^2}{t_\phi} 
  + y^4 C_{(2)}
\frac{s^2 (1-c_\theta)^2} {t_\phi^2} 
\ .
\end{eqnarray}
where $m \equiv m_t / \sqrt{s}$.  As odd powers of $c_\theta$ contribute to a
forward-backward asymmetry, we see that both the scalar exchange amplitude 
interfering with the SM amplitude as well as the scalar amplitude squared can
produce a nonzero asymmetry.  The resulting hadronic asymmetry will be
somewhat washed out by the PDFs, higher order jet radiation, and detector
effects.

To assess the impact of the scalar exchange on top observables at the Tevatron, we
compute the inclusive cross section and $A_{FB}^t$.  We do not attempt to
simulate the top decays, parton showering, hadronization, or detector effects, as these have
already been unfolded in the quoted measurements.  Our calculations are performed
using a version of MadGraph \cite{Cho:2006sx} (with CTEQ6L PDFs \cite{Nadolsky:2008zw} and
renormalization and factorization scales set to the top mass 
$\mu_R = \mu_F = m_t$ \cite{:2009ec})
which has been modified to insert by hand the exotic color factors and fermion-number violating
interactions of the form $\psi^T_i C^{-1} \psi_j \phi$.  We apply the SM $K$-factor of
$K = 1.329$ to account for higher order QCD corrections.  Since most of this 
$K$-factor arises from initial state radiation, we assume it will be similar for the new physics,
at least close to threshold.

\begin{figure}
\includegraphics[width=8cm]{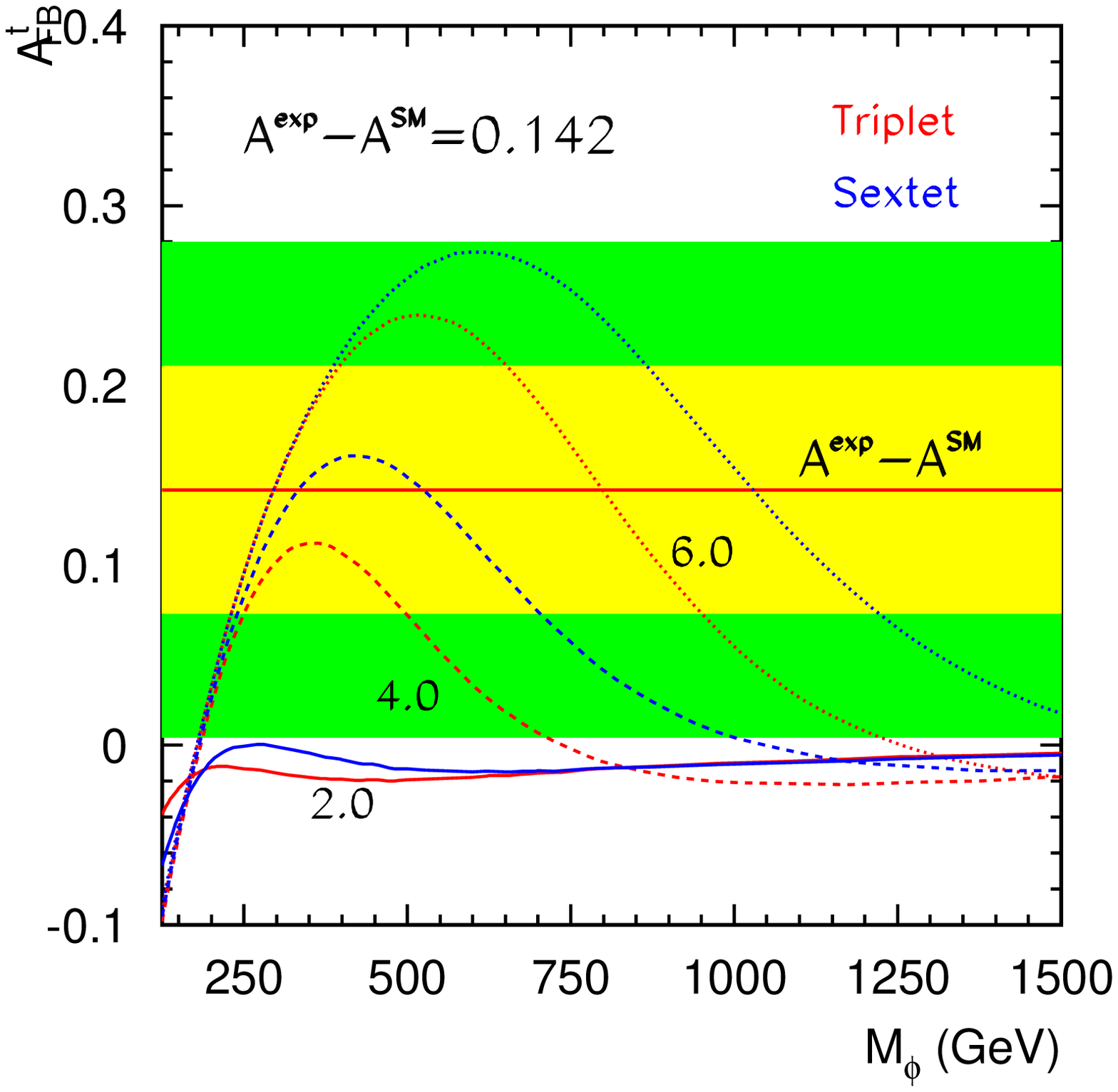}
\includegraphics[width=8cm]{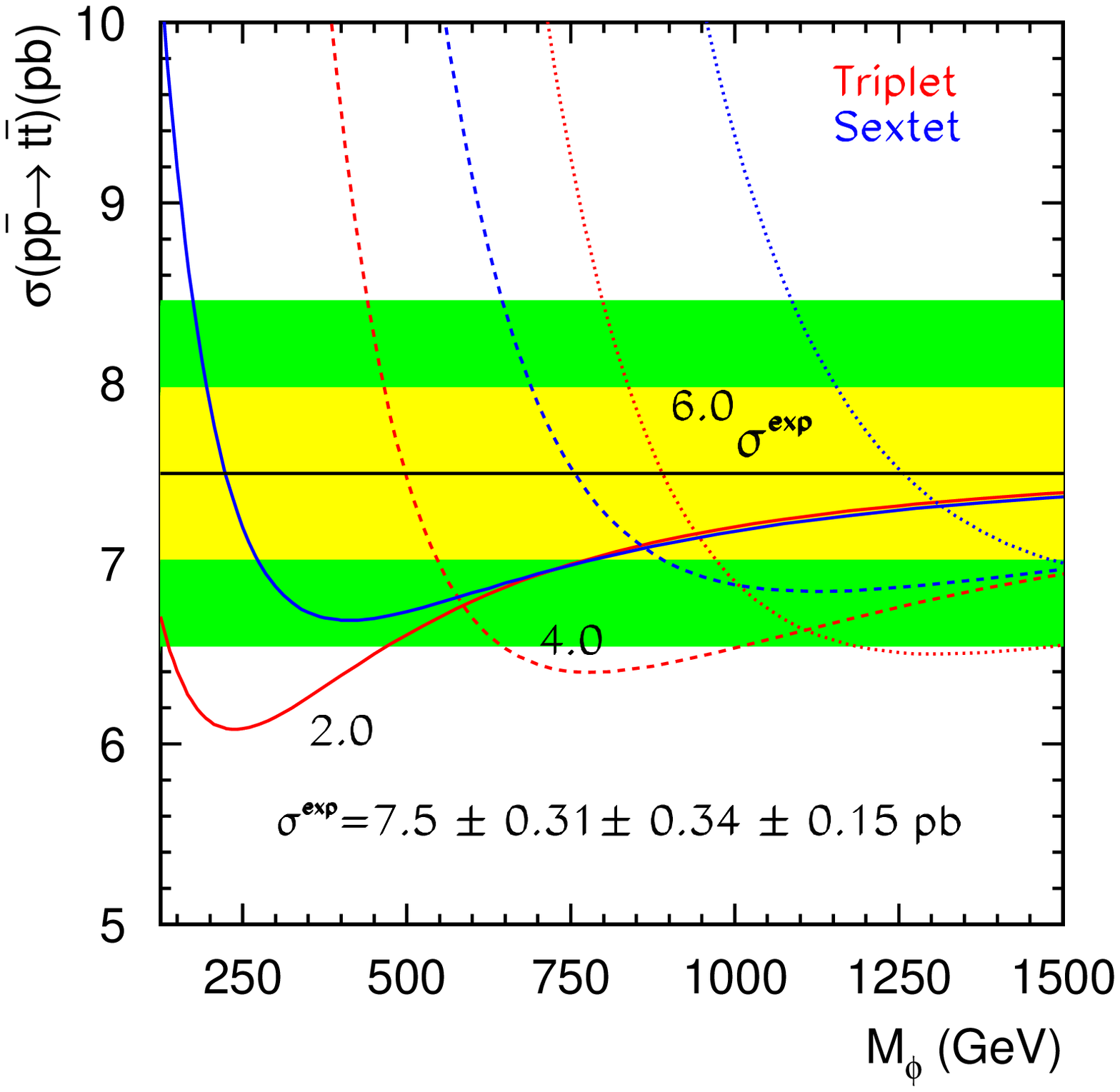}
\caption{\label{fig:afb36} Forward-backward asymmetry (left panel) and $t \bar{t}$ 
inclusive cross section (right panel) for color triplet (red curves) and sextet (blue curves)
scalars, as a function of the scalar mass $M_\phi$ for values of the 
coupling $y=2.0$ (solid curves),$y=4.0$ (dashed curves),
and $y=6.0$ (dotted curves).  Also shown are the range of values 
up to $1\sigma$ (yellow region) and $2\sigma$ (green region) from the experimental
measurements.}
\end{figure}

\begin{figure}
\includegraphics[width=8cm]{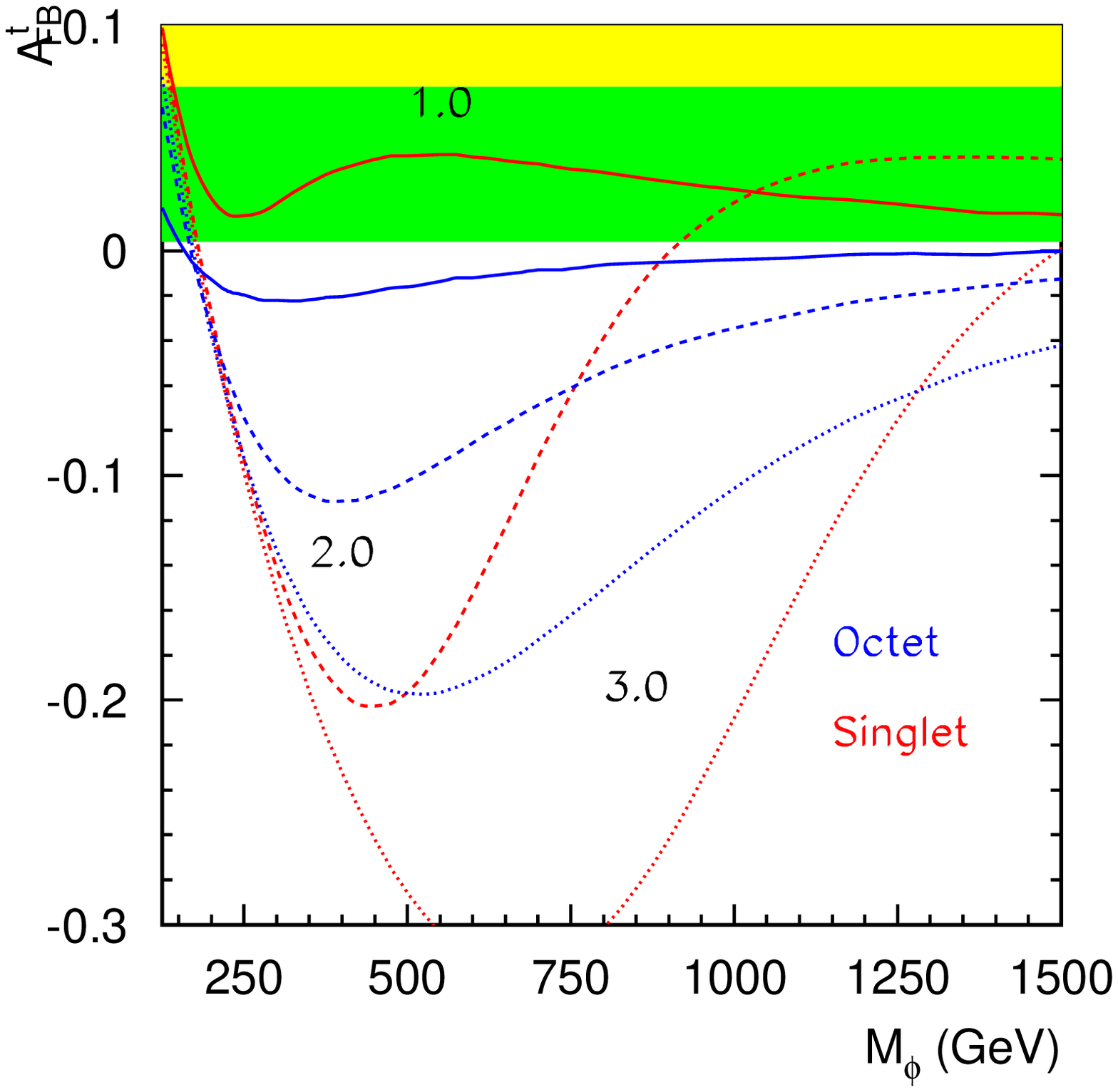}
\includegraphics[width=8cm]{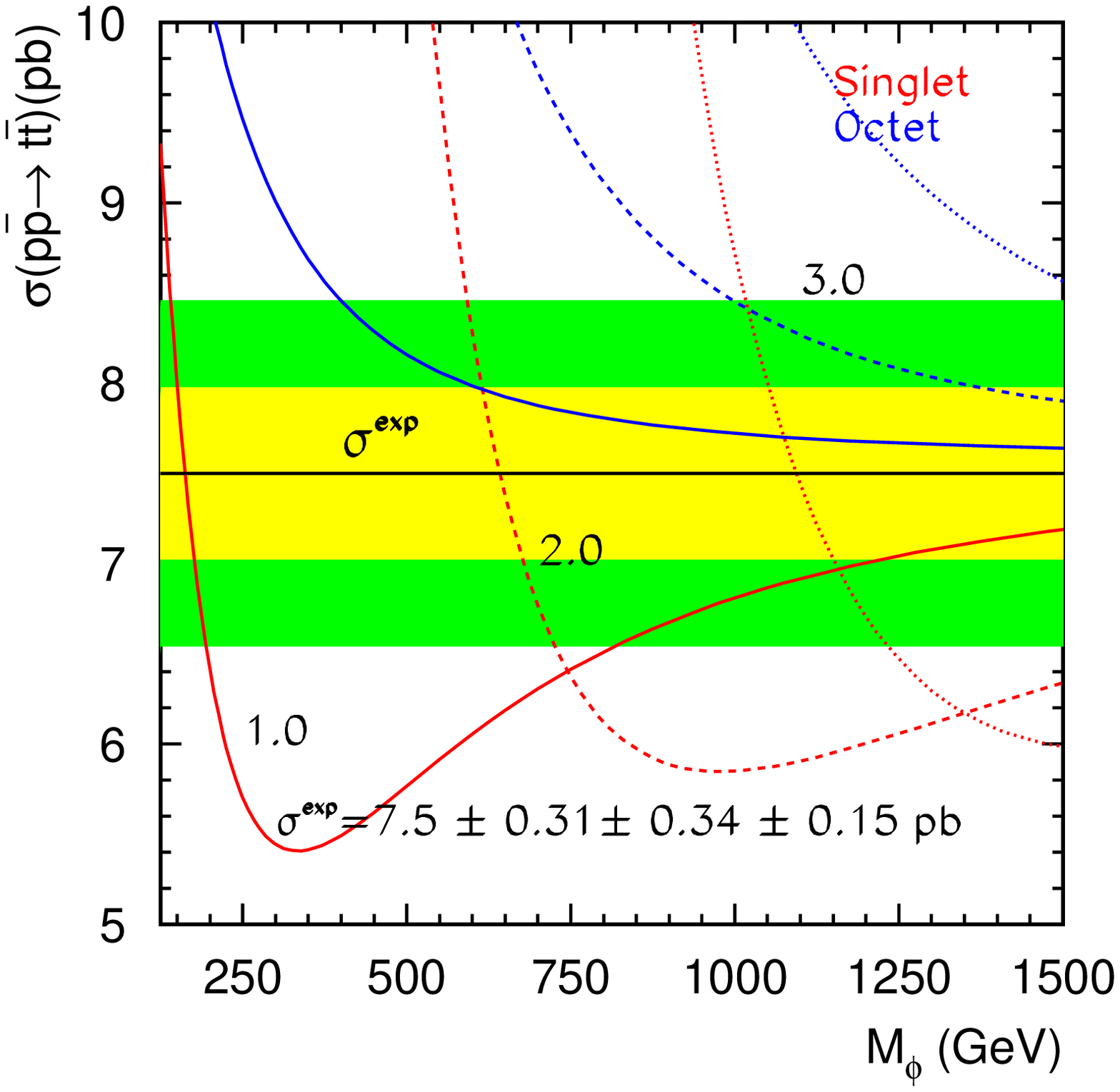}
\caption{\label{fig:afb18} Forward-backward asymmetry (left panel) and $t \bar{t}$ 
inclusive cross section (right panel) for color singlet (red curves) and octet (blue curves)
scalars, as a function of the scalar mass $M_\phi$ for values of the 
coupling $y=1.0$ (solid curves), $y=2.0$ (dashed curves)
and $y=3.0$ (dotted curves).  Also shown are the range of values 
up to $1\sigma$ (yellow region) and $2\sigma$ (green region) from the experimental
measurements.}
\end{figure}

\begin{figure}
\includegraphics[width=8cm]{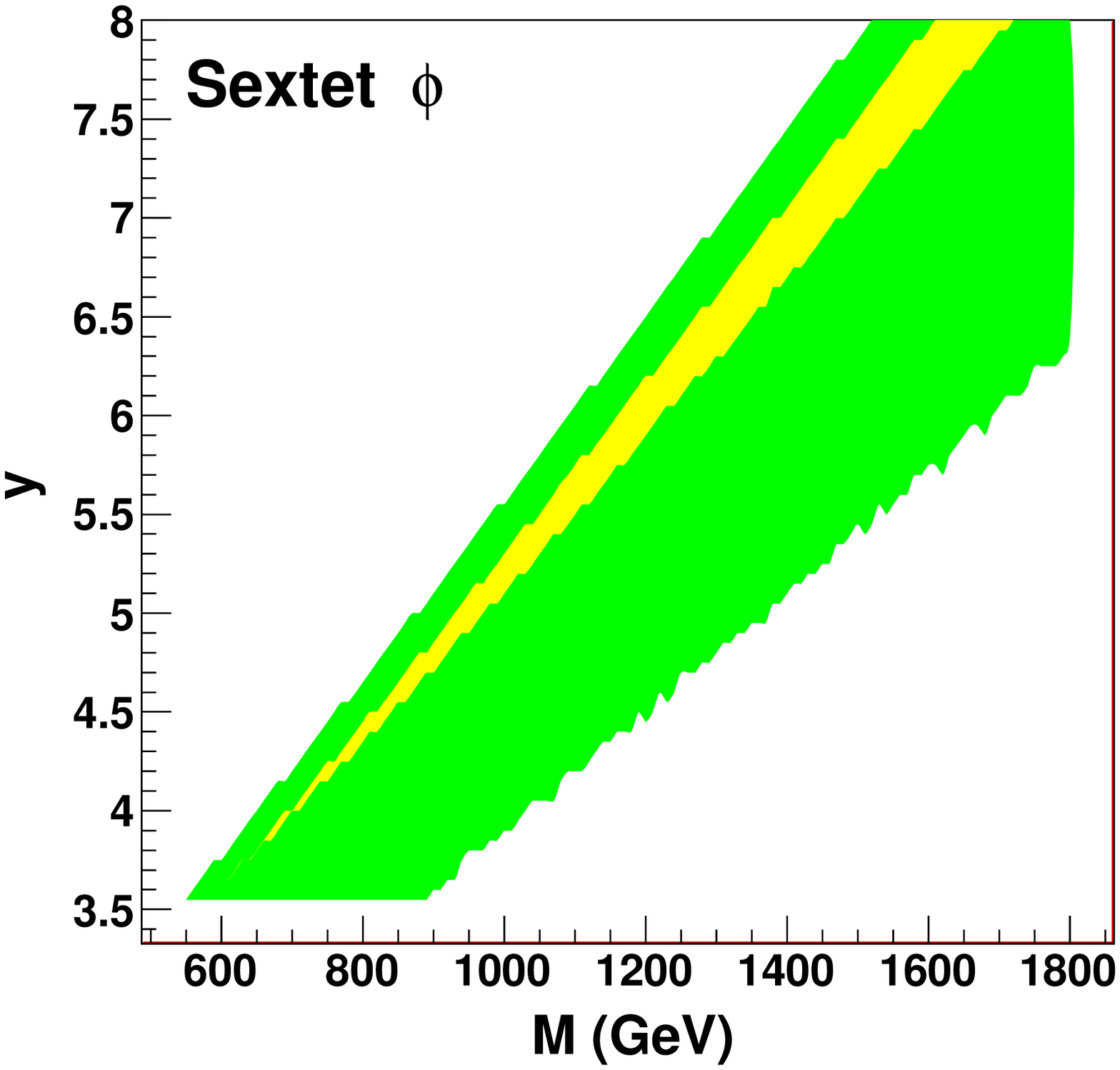}
\includegraphics[width=8cm]{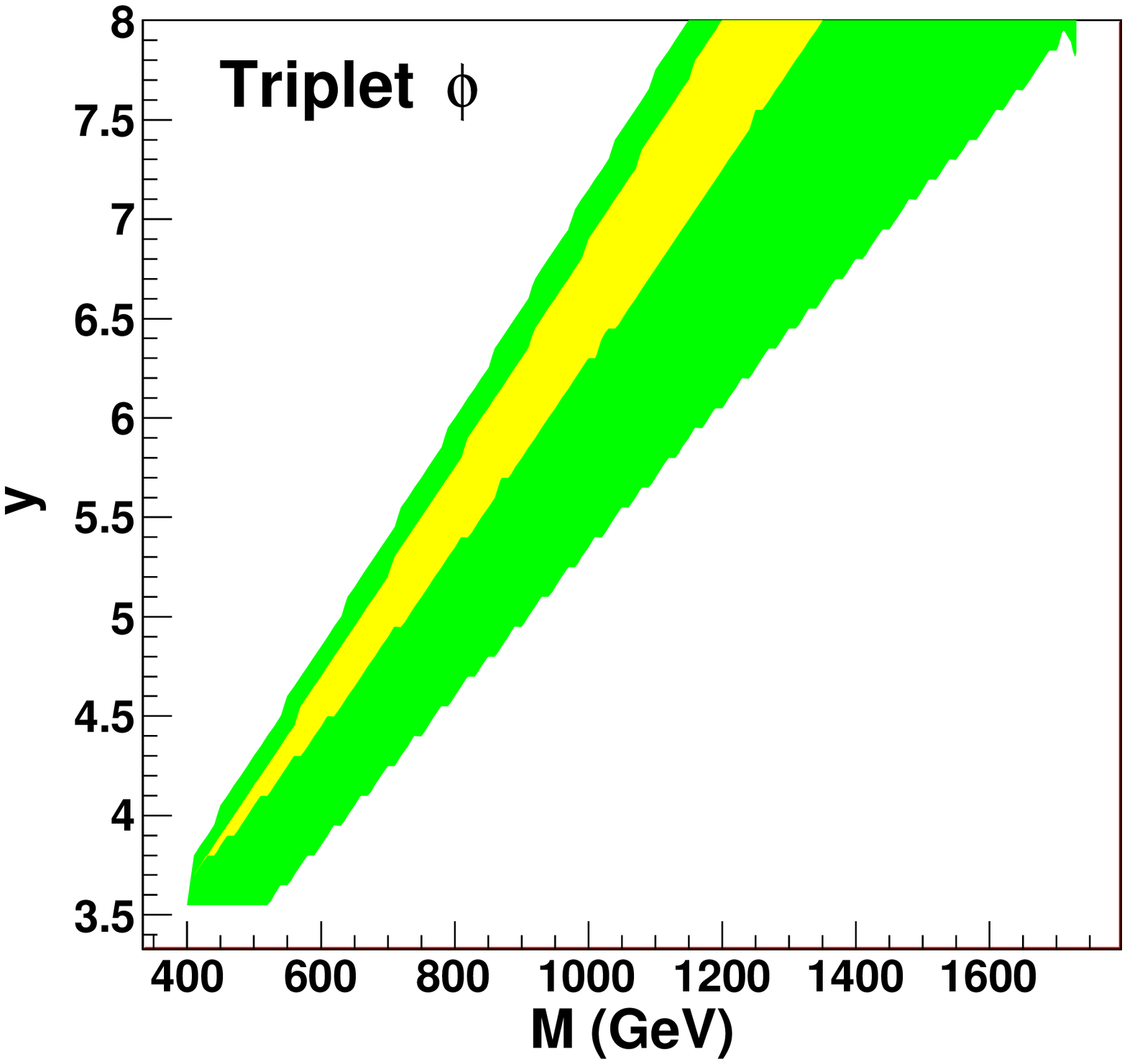}
\caption{\label{fig:scan} The regions of parameter space (in the $M_\phi$ - $y$ plane)
which predict  both $A_{FB}^t$ and $\sigma_{t \bar{t}}$
consistently within one sigma (yellow region) and
two sigma (green region) of the experimental measurements,
for color sextet (left panel) and color triplet (right panel) scalars. }
\end{figure}

In Figures~\ref{fig:afb36} and \ref{fig:afb18} we present the resulting $t \bar{t}$ cross section
and $A_{FB}^t$ for the four $\phi$ $SU(3)$ representations.  The experimental measurements
discussed above are shown for reference, where we have assumed the NLO QCD and
new physics contributions to $A_{FB}^t$ will simply add, since both are small corrections to the dominant tree level SM contribution.  
Examining the figures, it is clear that for both the color sextet and triplet cases, there are large parameter spaces for which the measurement of $A_{FB}^t$ can be explained to 
within $1\sigma$, and the cross section remains within the experimental errors.
In the intermediate region, $A_{FB}^t$ arises mainly from the numerator of the $t$-channel 
squared term so the color sextet and triplet have the desired sign. 
A large Yukawa coupling remains consistent with $\sigma_{t \bar{t}}$
because of a cancelation between the interference and scalar exchange 
squared terms.
A detailed scan of the consistent parameter space is shown in figure~\ref{fig:scan}.
In contrast, color octet and singlet
scalars both have great difficulty realizing a large positive contribution to $A_{FB}^t$, 
and only succeed for very low masses which have too large an impact on the cross section, and
open the door to further constraints by allowing the decay $t \rightarrow \phi u$ to take place.

\begin{figure}
\includegraphics[width=8cm]{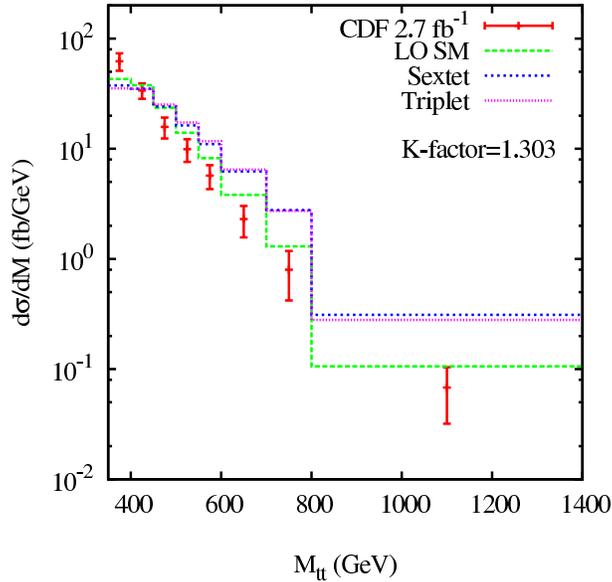}
\caption{\label{fig:mtt} $t \bar{t}$ invariant mass distribution for the SM, the theory with
a color sextet scalar, and the theory with a color triplet scalar.
The sextet ($M_{\phi}=610~\text{GeV}$, $y=3.65$) and triplet ($M_{\phi}=410~\text{GeV}$,$y=3.70$) 
benchmark points are within the 1$\sigma$ of both $A_{FB}^t$ and $\sigma_{t \bar{t}}$. }
\end{figure}

Another potentially important constraint results from the invariant mass distribution of the
$t \bar{t}$ pair. This distribution has been measured by CDF \cite{Aaltonen:2009iz} 
($m_t =175$ GeV) and is shown in Figure \ref{fig:mtt}.   Also shown is
the $t\bar{t}$ invariant mass distribution for benchmark points of the color sextet 
and triplet cases, and the SM predictions at the tree level with a constant K-factor. 
The modification of the invariant mass distributions for the $t$-channel scalar exchange 
shows similar behavior to the one for the vector boson exchange 
\cite{Jung:2009jz, Cheung:2009ch}, and there is tension between the predictions at the data.
However, a flat K-factor is probably overly naive, as the NLO enhancement arises
predominantly from processes involving extra radiation which are
suppressed at high invariant masses \cite{Frederix:2007gi}, which helps explain the
feature that the CDF data are consistently below the SM predictions for $M_{t\bar{t}} > 400$ GeV.  
Finally, since a genuine NLO calculation for the scalar exchange is beyond the scope of
this work, it is hard to derive solid constraints based on the invariant mass distributions.

\begin{figure}
\includegraphics[width=8cm]{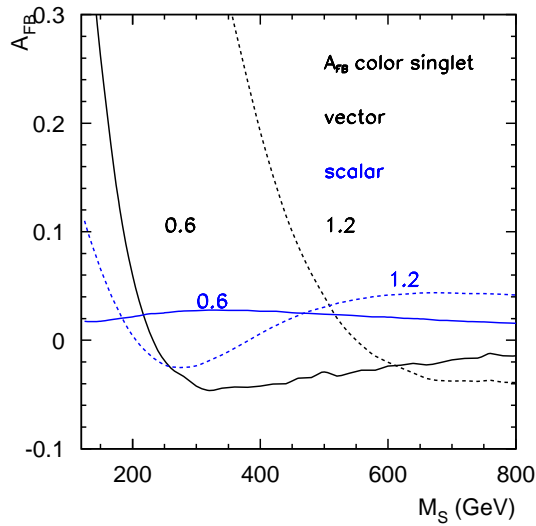}
\caption{\label{fig:compare} Comparison of $t$-channel excange from models with
vector bosons and scalars.}
\end{figure}

Compared to $t$-channel vector boson exchange, a light scalar produces a much smaller effect on $A_{FB}^t$ (see Figure \ref{fig:compare}).  This is because of
competition between the spin-correlation and the Rutherfold singularity for the 
scalar case, which can be seen from the opposite sign of the angular dependence between 
the numerator and denominator of the t-channel amplitude square term in 
Eq.~(\ref{eq: scalar_cos}).  One can understand this behavior intuitively (although for 
a massless top quark) by supposing an initial up quark moves in the positive $z$-direction,
scattering through a $t$-channel scalar particle to become a top quark.
As usual, the zero angular momentum partial wave has the largest component. 
However, in order to conserve angular momentum along the $z$-axis, 
the final top quark (which in the massless limit, has the opposite helicity state) 
must move in the negative $z$-direction, and consequently 
does not receive a Rutherford enhancement. 

\section{Constraints and Further Signals}

In this section, we discuss the constraints and future collider
signals for color sextet and triplet scalars.
Including the full $3 \times 3$ generational structure, the interactions are:
\bea
& & f_{ij}\overline{u^{c}}_{\alpha i} u_{\beta j} \phi_{\alpha\beta} \\
& & y_{ij}\epsilon_{\alpha\beta\gamma}\overline{u^{c}}_{\alpha i} u_{\beta j} 
\phi^{\prime}_{\gamma}~,
\eea
where $\phi_{\alpha\beta}$ is the color sextet scalar and 
$\phi^{\prime}_{\gamma}$ represents the color triplet scalar.
The symmetry of the color indices requires,
$f_{ij}=f_{ji}$, or $y_{ij}=-y_{ji}$.
Consequently, if the scalar is a color triplet, it cannot couple to the same flavor quarks.

Since we rely on large flavor violation in the colored scalar coupling to 
the up-type quarks, one may expect the model to suffer from 
serious constraints from $D^{0}-\overline{D^{0}}$ mixing and 
non-strange decays as $D\to \pi\pi$. 
However, these constraints limit the 
couplings to charm quarks \cite{Mohapatra:2007af,Chen:2009xjb},
and not the $\phi$-$u$-$t$ coupling which is of importance to explain $A_{FB}^t$.
Color sextet or triplet scalars are GIM-violating \cite{Arnold:2009ay}. Consequently, even with
suppressed couplings to charm, the left-handed rotation (which enters
the CKM matrix) can still induce flavor violation. 
This motivated our choice to explore the $(\bar{3},1)_{{4/3}}$ and $(6,1)_{4/3}$
representations (whose couplings depend on right-handed rotations, disconnected from
any SM quantity) in Eq.~(\ref{scalar}), as opposed to the
$(6,3)_{1/3}$ or $(\bar{3},3)_{1/3}$ representations, whose
couplings are connected to CKM elements.


A large $\phi$-$u$-$t$ coupling  could potentially contribute to the process
$uu \rightarrow tt$, strongly bounded by its production of same-sign dileptons
\cite{BarShalom:2008fq}.  $t$-channel exchange would bound the color singlet
and octet cases, but the sextet and triplets are also electrically charged, so $t$-channel
exchange does not lead to like-sign top quarks.  A color sextet could lead to this
process through $s$-channel exchange, provided $\phi$-$u$-$u$ and $\phi$-$t$-$t$
couplings are present and large enough to lead to an appreciable rate, but neither of
these couplings need be large for a significant contribution to $A_{FB}^t$.
 
 One can search for direct production of $\phi$, through partonic reactions such as,
\bea
u + g \to \bar{t} + \phi \\
\bar{u} + g \to t+\phi^{*}\\
g +g ~(q+\bar{q}) \to \phi^{*} + \phi~,
\eea
where the $\phi$ plus single top rates go through the Yukawa interactions $y$, whereas
the pair production rate is purely through the strong reaction, and thus depends only on
$M_\phi$ \cite{Chen:2008hh}.
If $\phi$ dominantly decays into a $ut$ final state, both processes will contribute to
$t \bar{t}$+jets.
From Fig.~\ref{fig:scan}, we see that to fit $A^{t}_{Fb}$ consistent with
$\sigma_{t\bar{t}}$, one can define benchmark models which
satisfy both measurements at $1\sigma$ by taking parameters which satisfy, 
\bea
\text{Sextet} &:&~y = {M_{\phi}\over 257~\text GeV}+1.28;\\
\text{Triplet} &:&~ y= {M_{\phi}\over 228.57~\text{GeV}}+1.8125 ~.
\label{benchmark}
\eea

\begin{figure}[th]
\includegraphics[width=8cm]{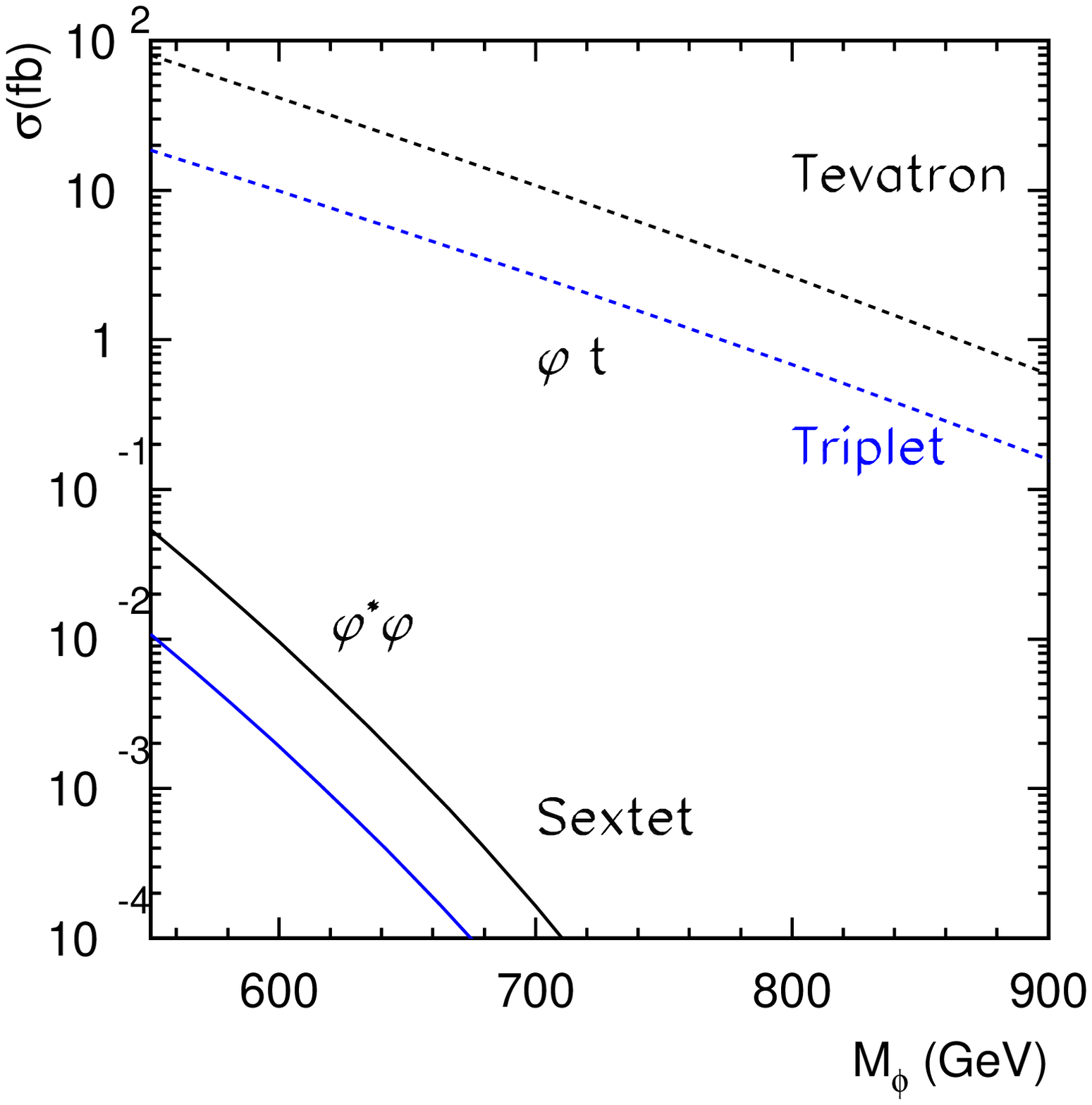}
\includegraphics[width=8cm]{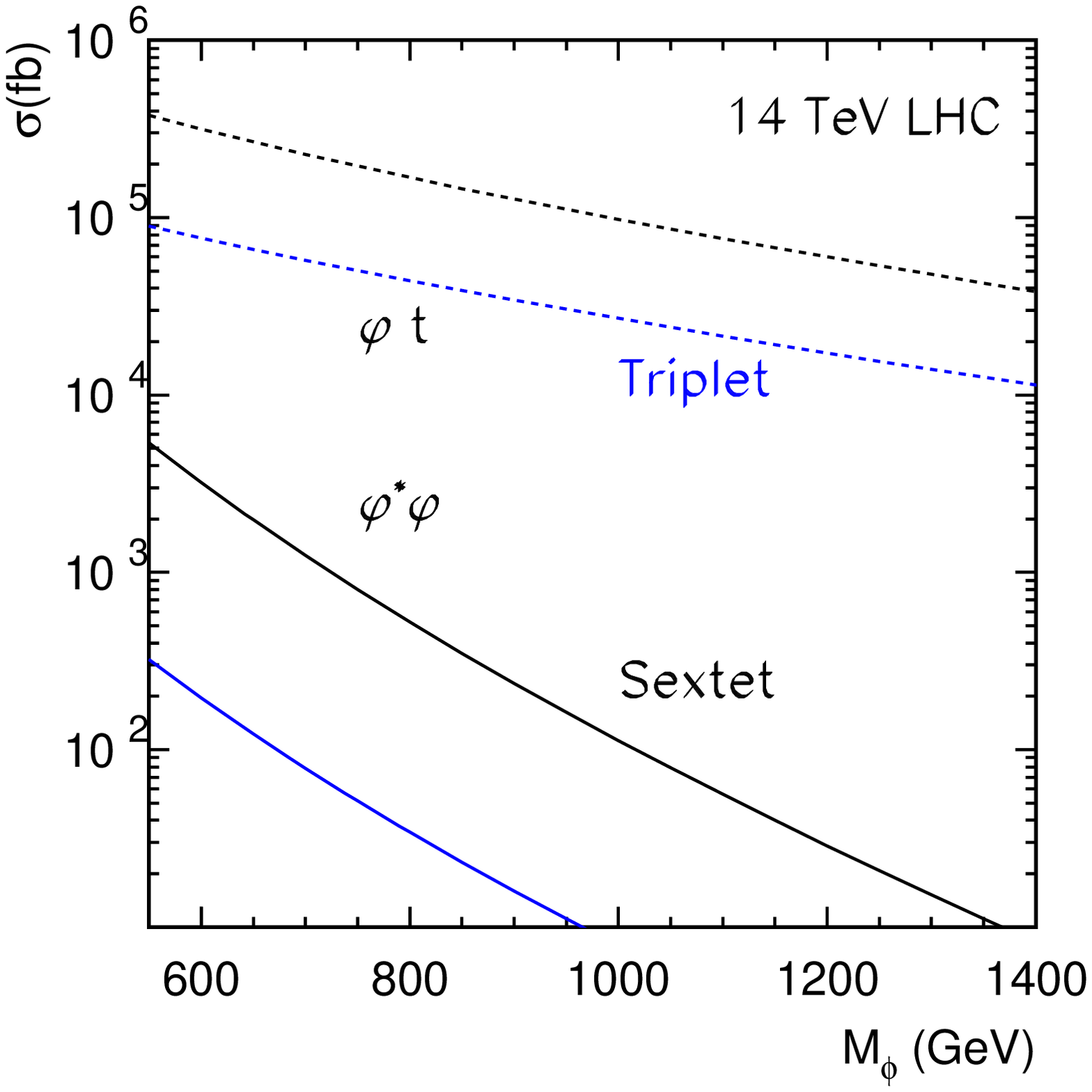}
\caption{\label{production} Production of triplet scalars (blue) and sextet scalars (black)
in the modes of
$\phi t$ (dashed curves) and 
$\phi\phi^*$ (solid curves) at Tevatron (a) and 14 TeV LHC (b)
for the benchmark models defined in the text. }
\end{figure}

In Fig. \ref{production}, we plot the rate of
$\phi \bar{t}+\phi^{*}t$ for the benchmark points described in Eq.~(\ref{benchmark})
as well as the $\phi^{*}\phi$ production rate at Tevatron and 14 TeV LHC. 
By assuming Br$(\phi\to u t)=100\%$, one can estimate the contribution to $t\bar{t}$+jets. 
At Tevatron, the contribution due to $\phi$ is less than 1\% of SM production due to 
phase space suppression.  At the 
14 TeV LHC, it can reach ${\cal O}(10^{2})$~pb which is comparable 
total SM $t\bar{t}$ rate (800 pb).
In the heavy resonace decay $\phi \to u t$, there is always a hard-jet associated with the top, and
this feature makes it possible to identify the events in the $t\bar{t}$+jets sample, for example
by taking the hardest jet and pairing it with a reconstructed top to see if there is any sign
of a resonance in such events.

\section{Outlook}

The forward-backward asymmetry of top quarks at the Tevatron is interesting, and bears watching.
While it is too early to conclude that it is a manifestation of new physics, it has consistently
been measured to be large for several years, and continues a general trend of mysterious behavior
in the measurements of heavy quark asymmetries.  Given the wealth of data we have about
the top quark, it is somewhat difficult to produce large effects while remaining consistent
with the other measurements.  Theories which manage that task thus reveal potential
holes in our knowledge of the dynamics of top, and inspire new searches to help us close them.

We have examined models in which a scalar is exchanged in the $t$-channel, exploring the
spectrum of $SU(3)$ representations which may explain the CDF measurement
of $A_{FB}^t$ while still allowing for a modest enough effect on the $t \bar{t}$ cross section to remain consistent with the experimental measurements.  The result of our exploration is
that color sextet and triplet scalars may be able to explain the anomaly, whereas color singlets
and octets cannot explain it without running into trouble with other observables.  The models that
work have relatively strong up-top flavor violating effects, and scalar particles with masses
in the range of 400 GeV to a bit less than 1.5 TeV.  They can lead to novel signatures that
affect the rates and kinematics of $t \bar{t}$ + jets, and warrant further attention as
the Tevatron collects data, and the LHC turns on.

\section*{Acknowledgements}
T Tait is glad to acknowledge conversations with fellow top enthusiast Michael Peskin,
and input and harassment from Tilman Plehn.
This work was supported by the World Premier International Research Center Initiative 
(WPI initiative) by MEXT, Japan. The work of J.S. was also supported by
the Grant-in-Aid for scientific research (Young Scientists (B)
21740169) from JSPS.  T Tait acknowledges the hospitality of the SLAC theory group, where
part of this work was completed.

\end{document}